\documentclass[aps,prl,preprintnumbers,reprint,superscriptaddress,raggedbottom]{revtex4-2}
\usepackage{graphicx}  % needed for figures
\usepackage{dcolumn}   % needed for some tables
\usepackage{bm}        % for math
\usepackage{epsfig,amsmath,amssymb,verbatim,mathrsfs,array,layout,textcomp,amssymb,latexsym,slashed}
\usepackage{xcolor}
\usepackage{siunitx}
\usepackage[colorlinks=true,citecolor=blue,urlcolor=blue,linktocpage=true,
linkcolor=blue]{hyperref}
\usepackage[utf8]{inputenc}
\usepackage{multirow}
\usepackage{amsthm}
\usepackage{thmtools}
\usepackage[capitalize,noabbrev]{cleveref}

\usepackage[normalem]{ulem}

\newcommand{\eV}{\mathrm{eV}}

\newcommand{\GeV}{\mathrm{GeV}}

\def\beq{\begin{equation}}
\def\eeq{\end{equation}}
\def\beqa{\begin{eqnarray}}
\def\eeqa{\end{eqnarray}}

\begin{document}

%%%%%%%%%%%%%%%%%%%%%%%%%%%%%%%
\title{Mass-Varying Dark Matter, Naturally}
 
\preprint{LAPTH-059/26}
%%%%%%%%%%%%%%%%%%%%%%%%%%%%%%%

%%%%%%%%%%%%%%%%%%%%%%%%%%%%%%%
%
\author{C\'edric Delaunay}
\email{cedric.delaunay@lapth.cnrs.fr}
\affiliation{Laboratoire d'Annecy de Physique Th\'eorique, CNRS--USMB, 9 chemin de Bellevue, 74940 Annecy, France}

\author{Seung J. Lee}
\email{sjjlee@kias.re.kr}
\affiliation{School of Physics, KIAS, 85 Hoegi-ro, Dongdaemun-gu, Seoul 02455, Korea}
\affiliation{Quantum Universe Center, KIAS, 85 Hoegi-ro, Dongdaemun-gu, Seoul 02455, Korea}

\author{Yuan Yin}
\email{yinyuan@kias.re.kr}
\affiliation{School of Physics, KIAS, 85 Hoegi-ro, Dongdaemun-gu, Seoul 02455, Korea}

\author{Bingrong Yu}
\email{bingrong.yu@cornell.edu}
\affiliation{Department of Physics, LEPP, Cornell University, Ithaca, NY 14853, USA}
%
%%%%%%%%%%%%%%%%%%%%%%%%%%%%%

%%%%%%%%%%%%%%%%%%%%%%%%%%%%%
\begin{abstract}
Ultralight scalars that vary dark-matter masses require protection from quantum corrections. Cyclic $\mathbb{Z}_N$ symmetries provide this protection but also suppress variation of the mean mass of equally populated partners. Unsuppressed variation typically requires asymmetric reheating through symmetry-breaking couplings that threaten radiative control. We show that scalar-induced mass splittings built into these constructions can instead bias relic abundances through symmetric interactions. The mechanism applies to thermal and nonthermal relics that reach internal equilibrium after production, and permits mass variation over a broad range of timescales.
\end{abstract}
%%%%%%%%%%%%%%%%%%%%%%%%%%%%%

\maketitle
%\tableofcontents

%\textit{Introduction.} 

Elementary particle masses need not be constant. The Brout--Englert--Higgs mechanism~\cite{Englert:1964et,Higgs:1964pj,Guralnik:1964eu} taught us that masses can arise from a background field. If that field evolves, masses can drift or oscillate, with diverse phenomenological consequences.  The relevant timescales range from slow evolution driven by scalars with masses near today's Hubble parameter $H_0\sim10^{-33}\,\mathrm{eV}$ to rapid oscillations of sub-eV scalars.

On cosmological timescales, scalar-dependent dark-matter (DM) masses exchange energy with the scalar, as in coupled-quintessence models~\cite{Amendola:1999er}. This exchange can slow the dilution of the DM energy density, yielding an older Universe~\cite{Anderson:1997un} or an effective phantom equation of state for dark energy (DE)~\cite{Das:2005yj}. Mass-varying neutrinos coupled to a light scalar may jointly drive cosmic acceleration~\cite{Fardon:2003eh}, while electron-mass variation near recombination may help resolve the Hubble tension~\cite{Baryakhtar:2024rky,Schoneberg:2026eys,Schoneberg:2026vaf}. Mass variation can also affect DM production and freeze-out~\cite{Heurtier:2019beu,Ferrante:2025avp}. 
Oscillating scalar backgrounds can also modulate Standard Model (SM) parameters, shifting atomic transition frequencies~\cite{Arvanitaki:2014faa,Stadnik:2015kia,Banerjee:2022sqg}, distorting neutrino oscillations~\cite{Krnjaic:2017zlz,Liao:2018byh,Losada:2022uvr}, or periodically varying the quark mixing matrix~\cite{Dine:2024bxv}.

The microscopic theories underlying these phenomena typically face a common naturalness problem. The interactions that generate mass variation also entail radiative corrections that destabilize the scalar potential, altering its dynamics~\cite{Doran:2002bc,DAmico:2016jbm,Marsh:2016ynw,Brzeminski:2020uhm}. A shift symmetry can protect the scalar potential~\cite{Goldstone:1962es}, but an appreciable variation of physical particle masses requires significant symmetry breaking that threatens the small curvature of the potential~\footnote{Super-renormalizable Higgs couplings provide a technically natural alternative~\cite{Piazza:2010ye,Graham:2015cka}.}.

Mirror~\cite{Delaunay:2025pho} and cyclic $\mathbb{Z}_N$ symmetries~\cite{Frieman:1995pm,Hook:2018jle,Brzeminski:2020uhm}, as well as monodromy~\cite{Bagherian:2026jxz}, can separate these effects. In $\mathbb{Z}_N$ constructions, a pseudo--Nambu--Goldstone boson (pNGB) couples to $N$ symmetry-related fermions. Individual physical masses vary linearly in the dimensionless shift-symmetry-breaking spurion $\epsilon$, whereas the field-dependent radiative potential first appears at order $\epsilon^N$.  Sufficiently large $N$ thus permits appreciable individual mass variations alongside a radiatively stable light scalar.

The response of a finite-density background, however, depends on the particle populations. For nonrelativistic particles, the energy density is governed by their population-weighted mean mass, whose scalar dependence can be much weaker than that of each physical mass. Equal populations of the $\mathbb{Z}_N$ partners suppress this collective response to $\mathcal{O}(\epsilon^N)$, as in the radiative potential. Unequal populations can evade this cancellation, but generating them through sector-dependent reheating couplings explicitly breaks $\mathbb{Z}_N$ and risks destabilizing the scalar potential.

Our key observation is that  relic populations need not respect the cyclic symmetry of the microscopic interactions. A homogeneous scalar background, for instance selected during inflation, \emph{spontaneously} breaks $\mathbb{Z}_N$ and splits the partner masses. Nonrelativistic relic formation can imprint this splitting onto the relative sector abundances, allowing an $\mathcal{O}(\epsilon)$ collective matter response without loss of radiative protection.

In this Letter, we develop a general framework for radiatively stable mass-varying DM, encompassing slowly rolling and oscillating scalars. We quantify the collective mass modulation and its dependence on conversion decoupling, showing how the same description applies to thermal and nonthermal relics that establish internal equilibrium after production. A companion paper~\cite{Delaunay:2026fse} applies this framework to interacting DM and scalar DE, where energy exchange can produce an apparent phantom crossing~\cite{Das:2005yj} motivated by DESI observations~\cite{DESI:2024mwx,DESI:2025fii,DESI:2025zgx}. We also discuss extensions to mass-varying neutrinos and electrons.\\

\textit{Modeling protected mass variation---}
To separate individual mass variation from radiative corrections, we consider a complex scalar $\Phi$ coupled to $N$ Dirac fermions, with canonical kinetic terms and mass interactions
\begin{equation}\label{eq:Lmass}
\mathcal{L}_\psi=-\sum_{k=0}^{N-1}
\left(M+y\,e^{2\pi ik/N}\Phi\right)
\bar\psi_{kL}\psi_{kR}+{\rm h.c.}\,.
\end{equation}
These preserve a cyclic $\mathbb{Z}_N$ symmetry under $\psi_k\to\psi_{k+1}$ and $\Phi\to e^{2\pi i/N}\Phi$, with indices understood modulo $N$. For $M=0$, they also preserve a continuous U(1) symmetry under $\Phi\to e^{i\alpha}\Phi$ and $\psi_{kR}\to e^{-i\alpha}\psi_{kR}$. The mass $M$ is the sole explicit U(1)-breaking spurion in this sector, formally transforming as $M\to e^{i\alpha}M$. Field redefinitions allow $M$ and $y$ to be taken real and nonnegative.

An invariant scalar potential spontaneously breaks U(1) at a scale $f$. Below the radial-mode mass, 
$\Phi=(f/\sqrt{2})e^{i\varphi/f}$ with $\varphi$ a pNGB. Defining $\theta_k\equiv\theta+2\pi k/N$, $\theta\equiv\varphi/f$, and $m\equiv yf/\sqrt{2}$, the physical fermion masses are
\begin{equation}\label{eq:psimass}
m_\psi(\theta_k)=
M\sqrt{1+\epsilon^2+2\epsilon\cos\theta_k}\,.
\end{equation}
The dimensionless spurion $\epsilon\equiv m/M\geqslant0$ controls their scalar dependence. For $\epsilon\ll1$, each mass therefore admits an $\mathcal O(\epsilon)$ fractional modulation over field excursions of order unity~\footnote{Interestingly, the cyclic fermion sector also admits a geometric realization with $\varphi$ as a Wilson-line scalar in a deconstructed five-dimensional U(1) gauge theory~\cite{Arkani-Hamed:2001kyx,Hill:2002kq,Arkani-Hamed:2003xts}, which will be developed elsewhere~\cite{Delaunay:2026uv}.}.

The fermion-induced vacuum potential instead sums over the full symmetry-related spectrum. At one loop, the Coleman--Weinberg (CW) potential is~\cite{Coleman:1973jx}
\begin{equation}
V_{\rm CW}(\theta)=
-\sum_k\frac{m_\psi^4(\theta_k)}{16\pi^2}
\left[\log\frac{m_\psi^2(\theta_k)}{\mu^2}-\frac{3}{2}\right].
\end{equation}
For $N>2$, its scalar-dependent part is finite and independent of the renormalization scale $\mu$~\cite{Frieman:1991tu,Frieman:1995pm}. The cyclic symmetry cancels all harmonics whose orders are not multiples of $N$~\cite{Hook:2018jle}, leaving, at leading order in $\epsilon$ and up to an additive constant,
\begin{equation}
V_{\rm CW}(\theta)\simeq
\frac{(-\epsilon)^N M^4}{4\pi^2(N-1)(N-2)}
\cos(N\theta)\,,
\end{equation}
whose curvature correction $\delta m_\varphi^2(\theta)\equiv f^{-2}\partial_\theta^2V_{\rm CW}$ begins at $\mathcal O(\epsilon^N)$ and is bounded by
\begin{equation}\label{eq:mphi}
\left|\delta m_\varphi^2(\theta)\right|
\lesssim
\frac{N^2}{(N-1)(N-2)}
\frac{\epsilon^N M^4}{4\pi^2f^2}\,.
\end{equation}

Radiative stability alone does not ensure that the available mass modulation is dynamically realized. Oscillations confined to one CW minimum explore only $\delta\theta\lesssim\mathcal O(1/N)$. An additional potential with a longer field-space period can support larger excursions. Such a contribution can arise from an anomalous breaking of the global $\mathbb{Z}_N$ symmetry by a confining gauge sector, generating, for example,
\begin{equation}\label{eq:Vnp}
V_{\rm np}(\theta)=
\chi_{\rm top}\left[1-\cos(\theta-\delta)\right]\,.
\end{equation}
Here $\chi_{\rm top}$ is the topological susceptibility of the confining sector in this cosine approximation, and $\delta$ is a phase offset relative to the CW contribution. The topological interactions underlying~\cref{eq:Vnp} preserve $\mathbb{Z}_N$ in perturbation theory. Provided the new sector introduces no other explicit breaking, the fermion-loop cancellation remains intact, as corrections that violate $\mathbb{Z}_N$ must vanish with this nonperturbative breaking.

The full vacuum potential is $V_0=V_{\rm CW}+V_{\rm np}$, with the nonperturbative term setting the bare mass squared $m_{\varphi}^2\equiv\chi_{\rm top}/f^2$ at its minimum. Radiative control requires $|\delta m_\varphi^2(\theta)|\ll m_{\varphi}^2$ throughout field space. For $N\gg1$, the CW correction becomes parametrically comparable to the bare mass squared near
\begin{equation}
N\simeq
\frac{2\log\!\big[M^2/(2\pi f m_{\varphi})\big]}
{\log(1/\epsilon)}\,,
\end{equation}
with each additional sector reducing the correction by approximately a factor $\epsilon$. As an example, $M=100\,\GeV$, $f=10^{10}\,\GeV$, $\epsilon=0.1$, and $m_{\varphi}=10^{-20}\,\eV$ give a maximum correction of about $3\%$ for $N=46$. 
With this hierarchy, the construction allows $\mathcal O(\epsilon)$ individual mass variation from initial displacement of order unity. The collective matter response further depends on the sector populations, as we discuss next.\\

%%%%%%%%%%%%%%%%%%%%%%%%%%%%%%%%%%%%%%%%%
\textit{Collective matter response---} 
In a dilute background of nonrelativistic $\psi$ fermions, the effective scalar potential is,
\begin{equation}
V_{\rm eff}(\theta)=V_0(\theta)+m_\psi^{\rm eff}(\theta)n_\psi\,,
\end{equation}
where $n_\psi\equiv\sum_k n_k$ is the total number density across the $N$ sectors, and
\begin{equation}
m_\psi^{\rm eff}(\theta)\equiv
\sum_{k=0}^{N-1}\eta_k m_\psi(\theta_k)\,,
\qquad
\eta_k\equiv\frac{n_k}{n_\psi}\,,
\end{equation}
is the population-averaged effective fermion mass. The effective potential governs scalar evolution through its derivative at fixed sector number densities. In a homogeneous Universe, the matter contribution can hold the scalar near a density-selected minimum, while its dilution can  allow the vacuum potential to drive further evolution and change the effective DM mass.

We characterize the collective matter response entering this dynamics by the rms fractional modulation
\begin{equation}
\Delta_\psi\equiv
\frac{\sqrt{\langle m_\psi^{\rm eff\,2}\rangle_\theta
-\langle m_\psi^{\rm eff}\rangle_\theta^2}}
{\langle m_\psi^{\rm eff}\rangle_\theta}\,,
\end{equation}
where $\langle X\rangle_\theta\equiv (2\pi)^{-1}\int_0^{2\pi} {\rm d}\theta X(\theta)$ denotes an angular average at fixed relative populations. Thus, $\Delta_\psi$ measures the mass modulation available across field space, while the variation realized in time depends on the scalar trajectory.
Although each physical fermion mass $m_\psi(\theta_k)$ depends on the value of the pNGB field at order $\epsilon$ (see~\cref{eq:psimass}), their average need not exhibit a comparable variation. Indeed, for a $\mathbb{Z}_N$-symmetric background with universal $\eta_k=1/N$, a shift $\theta\to\theta+2\pi/N$ merely permutes the equally weighted sectors and only harmonics that are multiple of $N\theta$ survive, yielding, $m_\psi^{\rm eff}(\theta)= N^{-1}\sum_k m_\psi(\theta_k)\equiv m_\psi^{\rm sym}(\theta)$ with
\begin{equation}
m_\psi^{\rm sym}(\theta)
\simeq M\left[
F(\epsilon)-\frac{(-\epsilon)^N}{\sqrt{\pi}N^{3/2}}\cos(N\theta)
\right]\,,
\end{equation}
where $F(\epsilon)\simeq 1+\mathcal{O}(\epsilon^2)$ is independent of $\theta$ and assuming the large $N$ limit. Thus, similarly to the CW potential, the matter response is strongly suppressed, with $\Delta_\psi^{\rm sym}\simeq \epsilon^N/\sqrt{2\pi N^3}$, despite the variation of the individual fermion masses.\\

% The same cancellation suppresses the backreaction of the fermion background on the scalar potential, leaving its slope and curvature effectively unchanged. This prevents the fermion background generating a dynamical misalignment through a density-dependent minimum. Nor can its subsequent dilution release a trapped field and thereby trigger a delayed onset of rolling. The scalar instead evolves under its vacuum potential and Hubble friction.

Evading this symmetry suppression requires a background that populates the sectors unequally.
For general population asymmetries $\delta\eta_k\equiv \eta_k -1/N$,
\begin{equation}
m_\psi^{\rm eff}(\theta)
= m_\psi^{\rm sym}(\theta)+ M\sum_{\ell\geqslant1}F_\ell(\epsilon)|\mathcal{A}_\ell|\cos\!\big(\ell\theta+\zeta_\ell\big)\,,
\label{eq:A1}
\end{equation}
where the sum runs over integers $\ell$ that are not multiples of $N$, 
\begin{equation}
\mathcal{A}_\ell\equiv\sum_{k=0}^{N-1}\delta\eta_k\,e^{2\pi i\ell k/N}\,,\quad \zeta_\ell\equiv \arg\mathcal{A}_{\ell}
\end{equation}
and $F_\ell(\epsilon)\equiv 2\sum_{j\geqslant0} C(1/2,j+\ell)C(1/2,j) \epsilon^{2j+\ell}$ with $C(n,k)$ denoting binomial coefficients. The $\mathcal A_\ell$ are Fourier moments of the population imbalance around the sector circle, measuring the incomplete cancellation of each mass harmonic. For illustration, consider a mixture of a symmetric population and a fraction $p$ concentrated in sector $k_\star$, $\delta\eta_k=p(\delta_{kk_\star}-1/N)$ where $0\leqslant p\leqslant1$. Then $\mathcal A_\ell=p\,e^{2\pi i\ell k_\star/N}$ for $\ell\notin N\mathbb Z$, giving
\begin{align}
    m_\psi^{\rm eff}(\theta)&\simeq  m_\psi^{\rm sym}(\theta)+\epsilon p M\cos\left(\theta+\frac{2\pi k_\star}{N}\right)\,,
\end{align}
and $\Delta_\psi\simeq[(\Delta^{\rm sym}_\psi)^2+\epsilon^2 p^2/2]^{1/2}$. 
Thus even a small population imbalance  $p\gg (\sqrt{2}/\epsilon)\Delta^{\rm sym}_\psi$ can overcome the symmetry suppression, and restores an appreciable $\mathcal{O}(\epsilon p)$ mass modulation for moderate $p$.\\

Asymmetric reheating can generate unequal fermion populations. If the DM states possess efficient interactions with a common SM bath, those can repopulate depleted $\mathbb{Z}_N$ partners and wash out the excess. Preserving it generally requires replicating the SM, coupling each partner to its own bath, and suppressing intersector energy transfer. Constraints on extra radiation then favor strongly asymmetric reheating, approaching the single-sector limit $p\to1$. Non-universal reheating couplings, however, explicitly break $\mathbb{Z}_N$. The resulting sector-dependent radiative corrections can spoil the cancellation of lower CW harmonics and reintroduce ultraviolet sensitivity in the scalar potential. The transmission of this breaking to the fermion-scalar sector must therefore be tightly controlled to preserve the shallow potential.\\
\begin{figure}[t]
\includegraphics[width=\columnwidth]{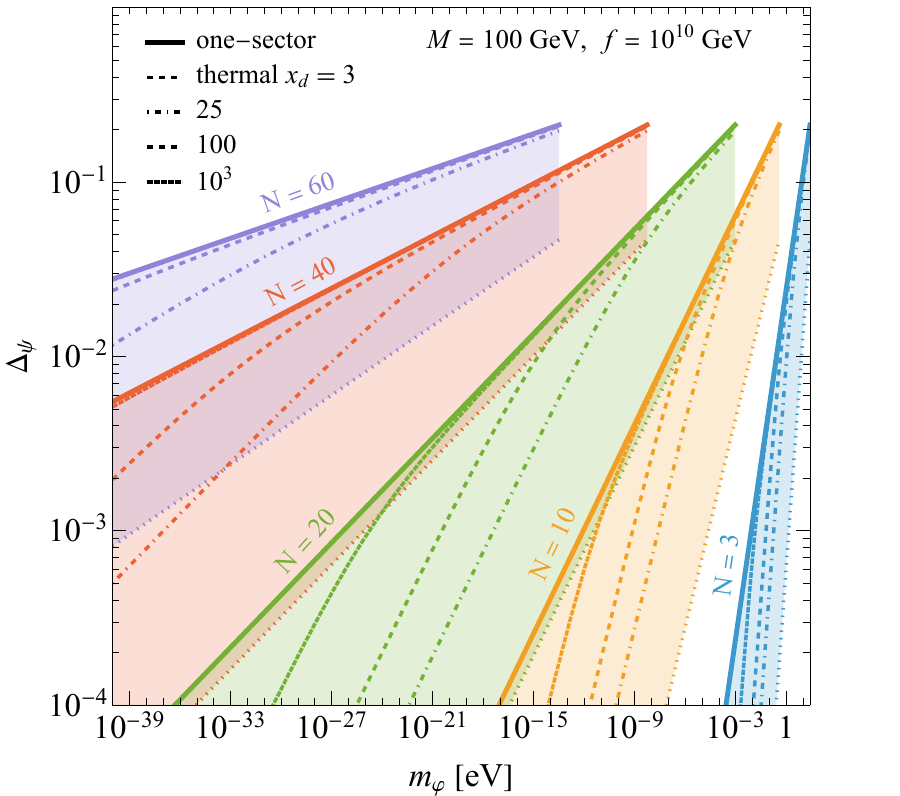}
\caption{Fractional fermion mass modulation $\Delta_\psi\simeq \epsilon|\mathcal{A}_1|/\sqrt{2}$ versus the minimal scalar mass $m_\varphi$ compatible with radiative control for the indicated values of $N$. Solid curves show the one-sector population ($p=1$) from asymmetric reheating; other curves show the asymmetric population resulting from thermalization for several values of $x_d\simeq {\rm min}[m_k]/T'_d$. The $m_\varphi$ axis scales as $(M^2/f)$.}
\label{fig:money-plot}
\end{figure}

\textit{Dynamical population imbalance---}
The required population asymmetry can instead arise dynamically from nonrelativistic relic production, with all DM partners coupled symmetrically to a single SM bath. We assume that $\theta$ remains approximately frozen during population formation. The $\mathbb{Z}_N$ cancellation suppresses leading thermal forces when the partners are relativistic and equally populated, while Hubble friction can suppress evolution driven by the nonperturbative potential provided $m_\varphi\ll H$ during this epoch. At a generic scalar background value $\theta_i$, for instance randomly selected by inflation, their different masses allow production and decoupling processes to distinguish them as they become nonrelativistic. Unequal relic populations can therefore emerge without introducing $\mathbb{Z}_N$-breaking interactions or additional SM copies, avoiding the reheating-induced corrections discussed above. 

The population imbalance generally depends on the specific relic-production mechanism. This dependence can nevertheless be erased if number-conserving intersector conversions establish internal thermal equilibrium and remain efficient after the total abundance is fixed {\it and} into the nonrelativistic regime. All partners then share a common temperature and chemical potential. The latter cancels from their relative abundances, giving~\cite{Griest:1990kh}
\begin{equation}\label{eq:etathermal}
\eta_k=\eta_k^{\rm eq}
\equiv
\frac{m_k^{3/2}e^{-m_k/T'}}
{\sum_jm_j^{3/2}e^{-m_j/T'}}\,,
\end{equation}
where $m_k\equiv m_\psi\!\left(\theta_i+2\pi k/N\right)$, even when the total density differs substantially from its zero-chemical-potential equilibrium value. Here $T'$ is the fermion temperature, equal to the SM temperature while kinetic equilibrium with the bath holds. The decoupling of conversions at a temperature $T_d'$ in the nonrelativistic regime freezes a mass-dependent population bias, provided subsequent redistribution and species-dependent depletion are negligible. The equilibrium form of the frozen fractions is therefore independent of the initial production mechanism, instead it is set by the fermion temperature and scalar background at conversion decoupling.  If these conditions fail, the abundances must instead be obtained from the coupled Boltzmann equations, as illustrated in studies of coannihilation without chemical equilibrium~\cite{Garny:2017rxs}.

For thermal freeze-out~\cite{Lee:1977ua,Steigman:1984ac}, the same interactions that allow the partners to annihilate into common SM bath particles can also mediate intersector conversions. These conversions can remain efficient after annihilation freeze-out, maintaining internal equilibrium while the total comoving abundance is conserved. For nonthermal production, such as freeze-in~\cite{McDonald:2001vt,Hall:2009bx}, sufficiently strong intersector interactions, for example mediated by a common dark radiation, can thermalize the fermions and equilibrate their relative populations after production, even if their interactions with the SM never establish thermal equilibrium~\cite{Chu:2011be}.

The mass modulation amplitude retains a memory of the scalar background at conversion decoupling. For small $\epsilon$ and large $N$, expanding the equilibrium weights in~\cref{eq:etathermal} while retaining their exponential bias gives
\begin{equation}\label{eq:meffthermal}
m_\psi^{\rm eff}(\theta)
\simeq
m^{\rm sym}_\psi(\theta)
-\epsilon M\frac{I_1(q)}{I_0(q)}
\cos(\theta-\theta_i)\,,
\end{equation}
where $I_0$ and $I_1$ are modified Bessel functions of the first kind, $q\equiv\epsilon\left(x_d-3/2\right)$ and $x_d\equiv {\rm min}[m_k]/T'_d$. The full derivation is given in the End Matter. For $q\ll1$, the modulation scales as $\Delta_\psi\propto\epsilon^2(x_d-3/2)$, while for $q\gtrsim1$, it becomes of order $\epsilon$. 

\Cref{fig:money-plot} shows the amplitude of the effective-mass dependence on the scalar across a broad range of ultralight scalar masses compatible with radiative control, $|\delta m_\varphi^2(\theta)|< m_\varphi^2$, for several values of $N$. The modulation and minimal scalar mass are linked through their common dependence on $\epsilon$, with the latter set by the CW potential. The physical scalar mass, set predominantly by the nonperturbative potential, can lie above it. Increasing $N$ allows substantial modulation to coexist with an extremely light, radiatively protected scalar. As conversion decoupling occurs deeper in the nonrelativistic regime, the thermal curves approach the single-sector population scenario, as for $\epsilon x_d\gg1$, Boltzmann suppression strongly favors the lighter partners. This shows that the modulation associated with maximally asymmetric reheating can be dynamically reproduced without replicating the SM or introducing additional $\mathbb{Z}_N$-breaking interactions.\\

%%%%%%%%%%%%%%%%%%%%%%%%%%%%%%%%%%%%%%%%%
\textit{Conclusions and outlook---} 
Therefore an appreciable scalar dependence of the population-averaged fermion mass can coexist with a radiatively protected ultralight scalar. The nonperturbative potential sets the scalar mass and permits large field excursions, while the cyclic symmetry suppresses the fermion-induced radiative correction. Unequal relic populations restore the collective matter response without introducing perturbative $\mathbb{Z}_N$-breaking interactions or SM replicas. If number-conserving intersector conversions maintain internal equilibrium after the total abundance is fixed, the relative populations are set by the fermion temperature and scalar background at conversion decoupling. 

The framework accommodates both thermal and nonthermal relic production, provided intersector conversions remain efficient after production and the scalar is approximately stationary while the relative populations are established. Once these populations freeze, the effective DM mass responds to subsequent scalar rolling or oscillations. Two regimes remain to be explored for either production mechanism. If the scalar oscillates during production, the evolving mass spectrum can affect the population bias. If conversions are inefficient during production and do not subsequently establish internal equilibrium, the final relative abundances retain a dependence on the production processes themselves.

Extensions to electrons and active neutrinos are challenging. First, additional chiral fermions must be introduced to cancel anomalies induced by the replicated chiral SM leptons. Moreover, light partners carrying SM electroweak charges face stringent bounds, for instance from precision measurement of the $Z$ width. Placing these partners in separate SM copies avoids these bounds but requires an asymmetric thermal history to suppress extra radiation. Additional explicit $\mathbb{Z}_N$ breaking introduced through sector-dependent reheating must then be sufficiently isolated from the fermion–scalar sector to preserve radiative protection.\\

%%%%%%%%%%%%%%%%%%%%%%%%%%%%%%%%%%%%%%%%%
\textit{Acknowledgment---}
We thank ChatGPT for useful comments on the manuscript. S.J.L. is partly supported by the Samsung Science and Technology Foundation under
Project Number SSTF-BA2201-06. BY was partly supported by the NSF grant PHY-2309456.

%%%%%%%%%%%%%%%%%%%%%%%%%%%%%%%
\bibliographystyle{apsrev4-2}
\bibliography{refs.bib}

%%%%%%%%%%%%%%%%%%%%%%%%%%%%%%%%%%%%
\clearpage
\onecolumngrid
\begin{center}
    \textbf{END MATTER}
\end{center}
\twocolumngrid

\appendix
\setcounter{secnumdepth}{1}
Here we derive the effective mass for thermally distributed partner populations and show how the large-$N$ limit simplifies its dependence on the scalar field.
Working at leading order in $\epsilon$, we retain the full
$\epsilon x_d$ dependence of the Boltzmann factor while
neglecting corrections of order $\epsilon^2x_d\ll1$. The abundance fractions in~\cref{eq:etathermal} then become 
\begin{equation}
    \eta_k^{\rm eq}\simeq \frac{1}{N Z_N(q,\theta_i)}e^{-q \cos(\theta_i+ 2\pi k/N)}\,,
\end{equation}
where
\begin{equation}
    Z_N(q,\theta_i)\equiv \frac{1}{N}\sum_{k=0}^{N-1}e^{-q\cos(\theta_i+2\pi k/N)}\,.
\end{equation}
After conversion decoupling, these frozen populations give
\begin{align}
    m_\psi^{\rm eff}(\theta)\simeq&\,m_\psi^{\rm sym}( \theta) -\epsilon M\bigg[ \frac{\partial \log Z_N}{\partial q} \cos(\theta-\theta_i)\nonumber\\
    &\, +\frac{1}{q}\,\frac{\partial \log Z_N}{\partial\theta_i} \sin(\theta-\theta_i)\bigg]\,.
\end{align}
Expanding the Boltzmann factor in modified Bessel functions
and performing the root-of-unity sum yields
\begin{equation}
Z_N(q,\theta_i)=I_0(q)+2\sum_{\ell\geqslant1} (-1)^{N\ell} I_{N\ell}(q)\cos(N\ell\theta_i)\,.   
\end{equation}
These higher-order terms encode the dependence of the frozen
populations on the formation angle $\theta_i$. All harmonics
of the evolving field $\theta$ whose orders are multiples
of $N$ remain confined to $m_\psi^{\rm sym}(\theta)$ and
cancel exactly in the population-dependent difference. At large $N$ limit  and fixed $q$, $I_{N\ell}(q)\sim(q/2)^{N\ell}/(N\ell)!\to0$, so that
$Z_N\to I_0(q)$ and
\begin{equation}
    m_\psi^{\rm eff}(\theta)\simeq
    m_\psi^{\rm sym}(\theta)
    -\epsilon M\frac{I_1(q)}{I_0(q)}
    \cos(\theta-\theta_i)\,.
\end{equation}

\end{document}